\RequirePackage{amsmath}
\documentclass[envcountsame,smallcondensed]{svjour3}

\usepackage[utf8]{inputenc}
\usepackage{fullpage}
\usepackage{xcolor}
\usepackage{siunitx}
\usepackage{lmodern}

\newcommand{\pdiff}[2]{\frac{\partial #1}{\partial #2}}
\newcommand{\cdiff}[2]{\frac{\mathrm{D} #1}{\mathrm{D} #2}}
\newcommand{\maxzero}[1]{\left[#1\right]^+}
\newcommand{\intd}[1]{~\mathrm{d}#1}
\newcommand{\iop}{\Delta P}
\newcommand{\fibredirection}{\psi}
\newcommand{\growingzone}{\Gamma}

\usepackage{amssymb,amsfonts,graphicx,bm}
\usepackage{pstricks,url,overpic,pst-plot,pst-node,rotating}
\usepackage[numbers,sort&compress]{natbib}

\makeatletter
\def\cl@chapter{\@elt {theorem}}
\makeatother
\usepackage{hyperref}
\usepackage[sort&compress,noabbrev]{cleveref}

\journalname{Journal of Elasticity}

\begin{document}

\title{A shell model of eye growth and elasticity}


\author{L. S. Kimpton  \and
        B. J. Walker \and
        C. L. Hall \and
        B. Bintu  \and
        D. Crosby  \and
        H. M. Byrne  \and
        A. Goriely}


\institute{L. S. Kimpton \and B. J. Walker \and H. M. Byrne \and A. Goriely
           \at
           Mathematical Institute, University of Oxford, Andrew Wiles Building, Radcliffe Observatory Quarter, Woodstock Road, Oxford, OX2 6GG, UK. \\
           Tel.: +44 (0)1865 273525\\
           \email{benjamin.walker@maths.ox.ac.uk}
           \and
           C. L. Hall \at
           Department of Engineering Mathematics, University of Bristol, Merchant Venturers Building, Woodland Road, Bristol, BS8 1UB 
           \and
           D. Crosby \at
           Wave Optics Ltd, 41 Park Drive, Milton Park, Abingdon, OX14 4SR, UK.
           \\
           Eyejusters Ltd, Unit 6, Curtis Industrial Estate, North Hinksey Lane, Oxford, OX2 0LX, UK.
           \and
           B. Bintu \at
           Department of Physics, Harvard University, Cambridge, Massachusetts 02138, USA.
}

\date{Received: date / Accepted: date}

\maketitle

\begin{abstract}
The eye grows during childhood to position the retina at the correct distance
behind the lens to enable focused vision, a process called emmetropization.
Animal studies have demonstrated that this growth process is dependent upon
visual stimuli, while genetic and environmental factors that affect the
likelihood of developing myopia have also been identified. The coupling
between growth, remodeling and elastic response in the eye is particularly
challenging to understand. To analyse this coupling, we develop a simple model
of an eye growing under intraocular pressure in response to visual stimuli.
Distinct to existing three-dimensional finite-element models of the eye, we
treat the sclera as a thin axisymmetric hyperelastic shell which undergoes
local growth in response to external stimulus. This simplified analytic model
provides a tractable framework in which to evaluate various emmetropization
hypotheses and understand different types of growth feedback, which we
exemplify by demonstrating that local growth laws are sufficient to tune the
global size and shape of the eye for focused vision across a range of
parameter values.
\end{abstract}

\section{Introduction}
\label{Intro}
In health, the human eye grows during childhood in order to adopt the correct
size and shape for focused vision in a process called
\textit{emmetropization}. The goal of emmetropization is to position the
retina at the correct axial distance behind the lens for the optical power of
the anterior eye. When this process fails, the individual is either myopic
(short-sighted, with excessive axial length) or hyperopic (long-sighted, with
insufficient axial length). Whilst, for many, myopia is readily treatable with
prescribed lenses, severe myopia (classified as a refractive error of ${-5}$
diopters or more, where 0 diopters signify normal vision) is correlated with
an elevated risk of secondary conditions, including neovascularization of the
retina, posterior staphylomas and macular holes, which can lead to blindness
\cite{Morgan2012}. Myopia is the most common cause of poor vision worldwide,
affecting almost one-third of the population \cite{Spillmann2020}, and
uncorrected refractive error is the second most common cause of blindness
\cite{Sherwin2013}. There is also substantial evidence of increasing global
incidence of myopia; in some regions as many as 80-90\% of high school leavers
were myopic at the beginning of the last decade \cite{Morgan2012}, with
overall incidence expected to continue to increase
\cite{Dolgin2015,Holden2015,Holden2016}. This is attributed to a number of
factors, ranging from a reduction in time spent outside to an increase in time
spent reading \cite{Morgan1975,Morgan2012}.

There are treatments that aim to slow the progression of myopia in children,
as discussed in a number of comprehensive reviews
\cite{Manjunath2014,Russo2014,Wu2019}, with the most effective being the use
of atropine eye-drops, whilst others include non-traditional corrective
lenses. The theory underpinning the prescription of these lenses, which
under-correct a myopic eye, assumes that myopic blur detected at the retina
prevents or slows subsequent axial growth, hence reducing myopia progression.
Building upon this assumption, we suggest that a necessary step towards
understanding and improving the treatment of myopia is to first understand the
process of emmetropization. Thus, as the primary goal of this work, we will
aim to develop a simple mechanical model of eye growth that can be used to
investigate hypotheses for emmetropization.

\begin{figure}[htb]
\centering
\vspace*{0.5cm}
\hspace*{1cm}
\begin{overpic}[width=0.5\textwidth]{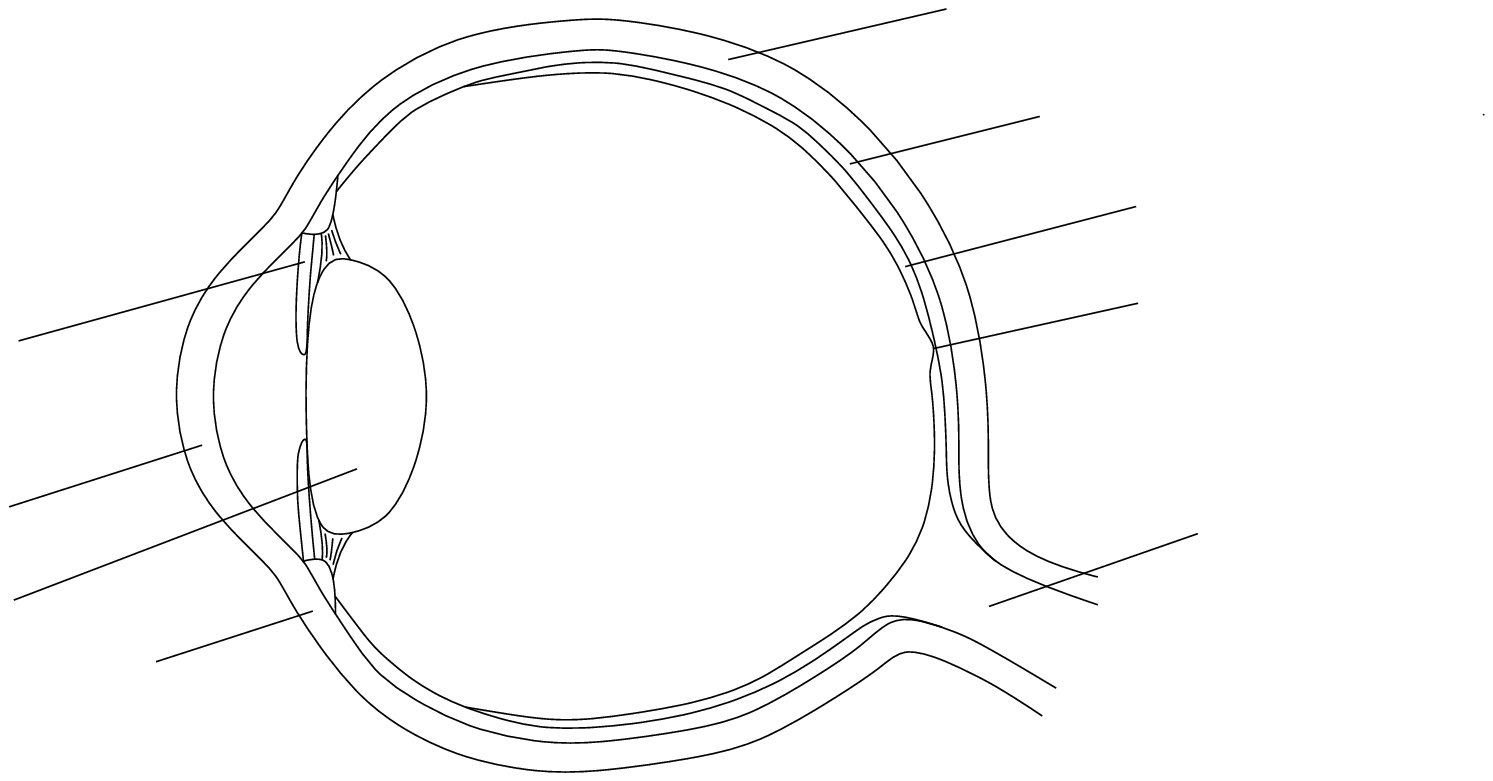}
 \put(65,50){\footnotesize Sclera}
 \put(71,44){\footnotesize Choroid}
 \put(78,38){\footnotesize Retina}
 \put(78,31){\footnotesize Fovea}
 \put(82,16){\footnotesize Optic nerve}
 \put(-9,28){\footnotesize Iris}
 \put(-17,17){\footnotesize Cornea}
 \put(-11,11){\footnotesize Lens}
 \put(-7,5){\footnotesize Limbus}
\end{overpic}
\vspace*{0.8cm}
\caption{A schematic diagram of a horizontal section through an eye, with many
of the structural features labelled. In this work, we will restrict attention
to explicit considerations of only the sclera, retina, and cornea, though we
will also utilise the optical properties of the anterior eye.}
\label{fig:eyeschem}
\end{figure}

The eye is an intricate organ, with many interconnected mechanical and optical
components. We illustrate some of these features in \cref{fig:eyeschem},
though in this work we will focus primarily on modeling the deformation and
growth of the sclera, upon which the retina and underlying choroid sit, with
minimal consideration of the anterior eye. The sclera contributes the majority
of the mechanical stiffness of the eye, with its thickness varying from
\SIrange{0.4}{1}{\milli\metre} \cite{Fatt1992}. Its mechanical properties
have been extensively modeled and measured \cite{Karimi2017,Romano2017}, with
sophisticated finite element models being employed to address the remodeling
typically associated with glaucoma and myopia
\cite{Grytz2012,Grytz2014,Grytz2017a}, with that of \cite{Grytz2017a}
explicitly modeling tree shrew sclera. High resolution models with similar
philosophies have also been applied to the mechanics of the cornea
\cite{Simonini2015,Sanchez2014}, including the recent patient-oriented study
of \cite{Pandolfi2020}. However, whilst the use of such techniques enables
systematic validation and fitting against clinical data, which have also been
particularly successful in developing the understanding of the growth of bones,
the heart, and arteries \cite{hu02,co04,gaogho06,kumahi07}, their
computational foundations also somewhat limit their scope for exploring these
complex systems.

Indeed, something not enjoyed by complex numerical models is a suitability for
analytical study and rapid exploration, a desirable trait of other models that
has been successfully exploited in similar contexts to identify plausible
growth laws and instabilities \cite{ta95,mogo11b,bustgo15}. These simpler
models, often described as toy models or caricatures, utilise simplified
geometries whilst retaining key properties of the full system, with the goal
not to obtain predictive models but to instead gain insight into the feedback
mechanisms between growth, remodeling, geometry and elasticity. This general
approach can be particularly valuable when consensus is lacking about the
underlying mechanisms, as is often the case in physiological systems. Thus, in
the context of emmetropization, there remains significant scope for a
simplified modeling framework that captures the mechanical and growth features
of the eye, with such a framework enabling future evaluation and exploration
of various hypotheses for the wide range of processes involved in ocular
development.

One such hypothesis concerns the drivers of scleral growth during
emmetropization. Supported by various observations in animals
\cite{Diether1997,Wallman1987,Foulds2013,Kroger1996,Liu2011,Werblin2007,Oyster1999},
though perhaps not directly translatable to the human eye, it is suggested
that at least part of the growth stimulus derives from locally interpreted
visual information on the retina, notably in combination with a host of
interacting genetic and environmental factors. Indeed, multiple works report
that the axial length of the grown eye in guinea pigs and chicks is dependent
on the wavelength of the incident light \cite{Foulds2013,Kroger1996,Liu2011},
suggesting a reliance of growth on the colour of observed light, which appears
consistent with the aforementioned incidence of myopia increasing in line with
reduced time spent outdoors. Here, utilising the optics of the anterior eye
and posing a simple model for growth stimulus that phenomenologically captures
this behaviour, we will consider this example hypothesis and test its ability
to produce qualitatively realistic morphologies.

We will proceed by formulating a simple but versatile model of
emmetropization, suitable for the exploration and evaluation of models of
growth and mechanical features of the developing eye. In particular, we will
focus on the later stages of emmetropization, in which the optical properties
of the anterior eye can be considered to be fixed, though this assumption may
be relaxed in future study. In response to visual stimuli, which we will later
define as the ability to correctly focus blue light at the retina, growth of
the scleral shell will act to relieve the local stress resulting from
intraocular pressure, which we explore as a possible driver for
emmetropization. We will also touch upon the effects of including
sophisticated mechanical properties, such as fibre reinforcement due to
collagen present in the sclera, as well as considering the effects of scleral
thickness so as to further showcase the tractability and flexibility of this
model approach.

\section{Model formulation}
\label{sec:formulation}
The sclera is modeled as a thin, morphoelastic shell that resists both bending
and stretching and is inflated by the intraocular pressure. Requiring the
shell to remain axisymmetric under loading and growth permits a choice of
coordinates that allows the model to be formulated solely in terms of
principal stretches, i.e. our representation of the deformation gradient is
diagonal. To describe the growth and the elastic response of the sclera, we
utilise the multiplicative decomposition approach formulated by
\cite{Rodriguez1994}, commonly referred to as \emph{morphoelasticity}. In this
formulation, the total deformation of the sclera in each principal direction
is written as the product of an elastic stretch and a `growth stretch'. This
has been applied in a wide range of biological applications, as reviewed by
\cite{Ambrosi2011} and \cite{Kuhl2014} and described recently by
\cite{Goriely2017}.

In our model, we suppose that the sclera deforms almost instantaneously in
response to changes in intraocular pressure, so that the elastic response is
very fast compared to growth, which occurs on the timescale of years.
Consequently, we assume mechanical equilibrium of the sclera at each instant
and determine the elastic stretches that characterise the deformation from the
unloaded, grown, reference configuration to the current, pressurised
configuration. The current position of the sclera is used to estimate the
degree of blur experienced locally at the retina, assuming smooth anterior
attachment to the cornea. This blur defines the growth stretches that are used
to update or `grow' the reference configuration at the next timestep. The
model formulation is detailed below and closely follows that used previously
to describe fungal growth and cell blebbing \cite{Tongen2006,Woolley2013}.

\subsection{Geometry}
\label{sec:geom}
\begin{figure}[htb]
\centering
\begin{overpic}[scale=0.7,tics=20]{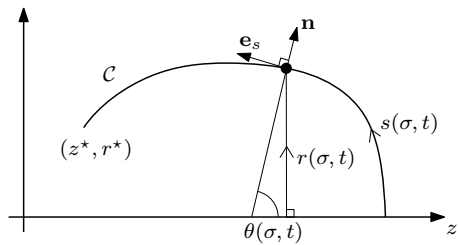}
 \put(100,0){\footnotesize $z$}
 \put(85,24){\footnotesize $s(\sigma,t)$}
 \put(66,16){\footnotesize $r(\sigma,t)$}
 \put(22,35){$\mathcal{C}$}
 \put(53,43){\footnotesize $\mathbf{e}_s$}
 \put(67,47){\footnotesize $\mathbf{n}$}
 \put(54,-1){\footnotesize $\theta(\sigma,t)$}
 \put(12,18){\footnotesize $(z^{\star},r^{\star})$}
\end{overpic}
\caption{Schematic diagram of the coordinate system in a plane of constant
$\phi$ for the deformed and grown scleral shell. The leftmost point is
$(z^{\star},r^{\star})$, the location of smooth attachment of the sclera to
the cornea.}
\label{fig:coord}
\end{figure}

We model the eye with a simplified geometry, with the centre of the pupil and
the fovea both lying on the anterior-posterior axis, about which the sclera is
axisymmetric. The physiological eye is not axisymmetric and the fovea sits
slightly temporal to the posterior pole, but axisymmetry is a reasonable
simplification. The position of the sclera is defined by rotating the curve
$\mathcal{C}$, which represents the centreline of the sclera about the
$z$-axis (see \cref{fig:coord}), with the retina lying on the interior surface
of the sclera.
The curve is parameterised by a material parameter $\sigma(\Sigma,t)$, which
is the arclength in the grown but unloaded configuration measured from where
$\mathcal{C}$ meets the $z$-axis at the rear of the eye. In turn,
$\Sigma\in[0,L]$ is an arclength parameter in the initial, unloaded
configuration, with smooth attachment to the cornea occurring at $\Sigma=L$.
Returning to the deformed configuration, $s(\sigma,t)$ is the arclength
distance from the $z$-axis, $r(\sigma,t)$ is the radial distance to the
$z$-axis and $\theta({\sigma},t)$ is the angle that the normal to
$\mathcal{C}$ makes with the $z$-axis, as shown in \cref{fig:coord}, so that
\begin{subequations}
\label{G1}
\begin{align}
 &\pdiff{r}{s}=\cos\theta\,,\label{G1a}\\
 &\pdiff{z}{s}=-\sin\theta\,.\label{G1b}
\end{align}
\end{subequations}
We associate a unit outward normal $\mathbf{e}_n$ and two unit tangent vectors
$\mathbf{e}_s$ and $\mathbf{e}_{\phi}$ with each point on the shell, which
point respectively in the direction of increasing $s$ and increasing $\phi$,
where $\phi$ is the angle that $\mathcal{C}$ is rotated about the $z$-axis.
These tangent vectors $\mathbf{e}_s$ and $\mathbf{e}_{\phi}$ are the principal
directions associated with the principal curvatures, $\kappa_s$ and
$\kappa_{\phi}$, where
\begin{subequations}
\label{G2}
\begin{align}
 &\kappa_s=\pdiff{\theta}{s}\,, \label{G2a}\\
 &\kappa_{\phi}=\frac{\sin\theta}{r}\,. \label{G2b} 
\end{align}
\end{subequations}
Finally, we denote the unloaded radius by $\rho(\sigma,t)$, analogous to
$r(\sigma,t)$ in the absence of elastic deformation, so that the elastic
stretches in the $\vec{e}_s,\vec{e}_{\phi},$ and $\vec{e}_n$ directions are
\begin{subequations}
\label{G3}
\begin{align}
 &\alpha_s(\sigma,t)=\pdiff{s}{\sigma}\,, \label{G3a} \\
 &\alpha_{\phi}(\sigma,t)=\frac{r}{\rho}\,, \label{G3b} \\
 &\alpha_n(\sigma,t)=\frac{h}{H}\,, \label{G3c}
\end{align}
\end{subequations}
respectively, where $h$ is the deformed scleral thickness and $H$ is the
unloaded scleral thickness.

\subsection{Force and moment balances}
Instantaneous mechanical equilibrium equations are obtained by balancing the
forces and torques acting on a small patch of the thin sclera as in
\cite{Woolley2013}. The sclera deforms in response to $\iop$, the difference
in intraocular and ambient pressure. We assume that both the intraocular and
ambient pressure remain constant, so that $\iop$ is set as constant throughout
the growth and remodeling process. The stress resultants $t_{\phi}$ and $t_s$
act along the shell in the $\mathbf{e}_{\phi}$ and $\mathbf{e}_s$ directions,
respectively, and both have units of force per unit length. The shear stress
resultant, which acts on surfaces with normal $\mathbf{e}_s$ in the direction
$\mathbf{e}_n$, is denoted by $q_s$ and has units of force per unit length. We
note that our assumption of axisymmetry ensures that no such shear acts on
surfaces with normal $\mathbf{e}_{\phi}$. The bending moments about the
$\mathbf{e}_{\phi}$ and $\mathbf{e}_s$ directions are written as $m_s$ and
$m_{\phi}$, respectively, and both have units of force. Neglecting inertial
effects, the momentum balance supplies
\begin{subequations}
\label{Mfull} 
\begin{align}
 \pdiff{}{s}\left(rq_s\right) &= r\left( \iop - \kappa_{\phi}t_{\phi} - \kappa_st_s \right)\,,\label{M1}\\
 \pdiff{}{s}\left(rt_s\right) &= \kappa_srq_s + t_{\phi}\cos\theta\,,\\
 \pdiff{}{s}\left(rm_s\right) &= m_{\phi}\cos\theta - rq_s\,.\label{M3}
\end{align}
\end{subequations}
A discussion of constitutive assumptions, used to define $t_s$, $t_{\phi}$,
$m_s$ and $m_{\phi}$ and close these mechanical equations, is postponed to
\cref{sec:mech_constitutive}, along with appropriate boundary conditions as
derived in \cref{sec:reduction}.

\subsection{Morphoelastic stretches and growth}
As outlined above, we assume a multiplicative decomposition of the
deformation, with the total deformation from the initial, unloaded
configuration to the grown, loaded configuration defined to be the product of
an elastic stretch and a `growth stretch'. Following the notation of
\cite{Goriely2017}, we denote the scalar total and growth stretches by
$\lambda_i$ and $\gamma_i$ ($i=s,\phi,n$), respectively, so that
\begin{subequations}
\begin{align}
 \lambda_s &= \alpha_s\gamma_s\,, \\
 \lambda_{\phi} &= \alpha_{\phi}\gamma_{\phi}\,, \\
 \lambda_n &= \alpha_n\gamma_n\,,
\end{align}
\end{subequations}
where the $\alpha_i$ $(i=s,\phi,n)$ are the purely elastic stretches defined
in \cref{G3}.

As defined in \cref{sec:geom}, $\sigma$ and $\rho$ denote the arclength and
radial distance in the unloaded state. That is, $\sigma$ and $\rho$ refer to
the `virtual configuration' that includes the part of the deformation due to
growth, but not the elastic response to the applied load. Recalling $\Sigma$
as the arclength in the initial, ungrown and unloaded configuration, and
denoting the associated radial distance by $R$, then the total and growth
stretches in the $\vec{e}_s$ and $\vec{e}_{\phi}$ directions may be expressed
simply as
\begin{subequations}
\begin{align}
 \lambda_s&=\pdiff{s}{\Sigma}\,, \\
 \gamma_s&=\pdiff{\sigma}{\Sigma}\,, \\
 \lambda_{\phi}&=\frac{r}{R}\,, \\
 \gamma_{\phi}&=\frac{\rho}{R}\,.
\end{align}
\end{subequations}
We suppose that each material point in the sclera has some capacity for growth
that could depend on a range of factors including age, position, and visual
stimuli. We also assume that each region of the sclera can increase in size
but cannot decrease, so that $\gamma_i
\geq 1$ ($i=s,\phi,n$). It is known that the sclera becomes thinner during
axial elongation, so, due to an absence of known mechanism, here we prohibit
growth in scleral thickness and set $\gamma_n=1$, though alternative routes
such as imposed mass conservation may easily be accommodated in this
framework. Thus, we have $\lambda_n = \alpha_n = h/H$.

In modeling the drivers of scleral growth, we are motivated by early
experimental studies in which intraocular pressure is associated with axial
elongation and related myopia in chick embyros \cite{Coulombre1956}, rabbits
\cite{Maurice1966}, and humans \cite{Quinn1995}, though the latter remains
controversial. Hence, we assume that regions of the sclera grow in response to
elastic stress. Seeking a minimal phenomenological model, we assume that
growth occurs to relieve the local strain, writing
\begin{subequations}
\label{Lagrange}
\begin{align}
 \cdiff{\gamma_s}{t} &= \eta\maxzero{\lambda_s -\gamma_s} = \eta\gamma_s \maxzero{\alpha_s - 1} \label{Lagrangea}\,,\\
 \cdiff{\gamma_{\phi}}{t} &= \eta\maxzero{\lambda_{\phi} -\gamma_{\phi}} = \eta\gamma_{\phi} \maxzero{\alpha_{\phi} - 1}\,, \label{Lagrangeb}
\end{align}
\end{subequations}
where $\maxzero{x}=\max(x,0)$ and $\eta$ denotes a growth rate that is here
assumed to be independent of direction, though may be generalised in future
work. In \cref{Lagrange}, `$\mathrm{D}/\mathrm{D}t$' denotes a material
derivative, holding $\Sigma$, the arclength in the original configuration,
fixed. These equations express that the relative rate of growth, $
\dot{\gamma} /\gamma$, is driven by the elastic strains in the system.
Since the intraocular pressure is held constant, with the system therefore
always loaded, equilibrium is only achieved when $\eta(\sigma,t)=0$, which we
discuss later when specifying this growth rate.

Since we anticipate that the final grown configuration will be far from the
initial reference configurations, it is both preferable and computationally
advantageous to express the dynamics in terms of the grown, unloaded
configuration. Holding $\Sigma$ constant, we write \cref{Lagrangea} as
\begin{equation}
 \pdiff{}{\Sigma}\left(\cdiff{\sigma}{t} \right) = \eta\maxzero{\pdiff{s}{\Sigma} - \pdiff{\sigma}{\Sigma}}\,.
\end{equation}
Then, after multiplying through by $\partial{\Sigma}/\partial{\sigma}$,
applying the chain rule and integrating, we find
\begin{equation}
 \cdiff{\sigma}{t} = \int_0^\sigma \eta\maxzero{\pdiff{s}{\hat{\sigma}} - 1} \intd{\hat{\sigma}}\,,\label{Grs}
\end{equation}
noting that $\partial{\Sigma}/\partial{\sigma}\geq0$. As $R$ is constant
whilst $\Sigma$ is constant, \cref{Lagrangeb} simply becomes
\begin{equation}
  \cdiff{\rho}{t} = \eta\maxzero{r - \rho}\,,\label{Grp}
\end{equation}
as $\partial R/\partial\rho\geq0$. If $\eta$ is constant, we recover the
growth laws described in \cite{Woolley2013}.

\subsection{Mechanical constitutive assumptions}
\label{sec:mech_constitutive}
The timescale for scleral growth is on the order of years, so the equations of
elastic equilibrium decouple from the growth laws and we consider \cref{G1},
\cref{G2a}, \cref{G3a} and \cref{Mfull} as a set of seven ordinary
differential equations in the eleven unknowns $r$, $z$, $s$, $\theta$,
$\kappa_s$, $\alpha_s$, $q_s$, $t_s$, $t_{\phi}$, $m_s$ and $m_{\phi}$, which
each depend on $\sigma$. To close this sub-problem, we specify a constitutive
relation for each of $m_s$, $m_{\phi}$, $t_s$ and $t_{\phi}$, thereafter
viewing \cref{G1,G2a,G3a,Mfull} as ordinary differential equations for $r$,
$z$, $\theta$, $s$, $q_s$, $\alpha_s$ and $\kappa_s$.

Firstly, we suppose that the curvature of the shell generates bending moments
according to
\begin{equation}
 m_s = m_{\phi} = E_B\left(\kappa_s + \kappa_{\phi} \right),\label{Cm1}
\end{equation}
where the bending modulus, $E_B$, has units of force multiplied by length.
This standard form of constitutive law is also used in
\cite{Tongen2006,Woolley2013}, though with a reference curvature. Here, we
omit such a reference curvature, which, if constant, would have no impact on
our model system, as we will see in \cref{sec:reduction}.

Next, we view the sclera as an incompressible, hyperelastic, fibre-reinforced
shell, accordingly assuming that the stress components $t_s$ and $t_{\phi}$
are functions of the principal elastic stretches $\alpha_s$ and
$\alpha_{\phi}$. A fibre-reinforced formulation allows us to model the
sclera's ability to resist tension, as well as incorporating anisotropic
effects arising from fibre orientation, consistent with the substantial amount
of collagen present in the sclera and the observations of
\cite{Girard2011,Grytz2014}.

Fibre modeling in soft tissues is particularly challenging. When considered as
a continuum, collagenous soft-tissues show a certain amount of fibre
distribution at each point \cite{jogiwh15,gogiet12,copije15}. In principle,
this angular distribution needs to be integrated at each point to obtain its
overall mechanical contribution. However, when the distribution is
sufficiently localized around different angles, the mechanical contribution of
fibres can be modeled using a finite number of reinforcing fibres \cite{sp72}.
The effect of a small fibre dispersion around these particular angles can also
be taken into account, amounting to a renormalization of the elastic constants
for both the isotropic and anisotropic parameters \cite{hoog10,medago15}.

Following our modeling philosophy of developing a tractable framework that
captures the relevant mechanical effects, we therefore model the anisotropic
response of the sclera by introducing two families of fibres. However, since
deformations have been assumed to be axisymmetric, the two fibre families must
be equal in strength and opposite in alignment with respect to the main axes.
Specifically, the two families of fibres  make an angle $\fibredirection$ with
$\vec{e}_{\phi}$, where the angle $\fibredirection$ is a function of position.
Denoting the fibre directions as $\vec{a}_0$ and $\vec{b}_0$ in the initial
reference configuration, we explicitly take
\begin{subequations}
\begin{align}
 \vec{a}_0&=\sin\fibredirection(\Sigma)\vec{e}_s + \cos\fibredirection(\Sigma)\vec{e}_{\phi}\,, \label{eq:fibre:a}\\
 \vec{b}_0&= \sin\fibredirection(\Sigma)\vec{e}_s - \cos\fibredirection(\Sigma)\vec{e}_{\phi}\,. \label{eq:fibre:b}
\end{align}
\end{subequations}
This reduces to purely circumferential reinforcement when $\fibredirection=0$,
in line with the observations of \cite{zhaljo15,copije15}, with dispersion
about this configuration being modeled by small values of $\fibredirection$.
Inherent to this general setup is the issue of isotropy at the base of the
sclera, where the surface intersects with its axis at $\Sigma=\sigma=s=0$,
which we resolve here by prescribing that $\fibredirection\to\pi/4$ as
$\sigma\to0$, though this may also be addressed by omitting a small region of
reinforcement close to this apex. Note that, here and throughout, fibre
orientation is treated as a material property, with
\begin{equation}
 \fibredirection(\sigma(\Sigma,t),t) = \fibredirection(\Sigma)\,, \label{matfibre}
\end{equation}
which we will later specify based on the observations of
\cite{Girard2011,Grytz2014}. We similarly treat $H$, the undeformed shell
thickness, as a material property, in particular following \cite{Fatt1992},
though we will also consider shells of initially uniform thickness for
comparison.

We now relate stress to strain via a strain-energy density, $W$, which, for
simplicity, depends only on the first invariant of the isotropic strain,
$I_1$, and the fibre stretch, $I_4$:
\begin{subequations}\label{Cpst}
\begin{align}
 I_1&=\alpha_s^2+\alpha_{\phi}^2+\alpha_n^2\,, \\
 I_4&=\alpha_s^2\sin^2\fibredirection+\alpha_{\phi}^2\cos^2\fibredirection\,.
\end{align}
\end{subequations}
The principal stresses are then given by
\begin{subequations}
  \begin{equation}
    t_s = 2H\alpha_n\left[\left(\alpha_s^2-\alpha_n^2\right)\pdiff{W}{I_1} + 2\alpha_s^2\sin^2\fibredirection\pdiff{W}{I_4} \right]\,, \label{Cpsts}
  \end{equation}
  \begin{equation}
    t_{\phi} = 2H\alpha_n\left[\left(\alpha_{\phi}^2-\alpha_n^2\right)\pdiff{W}{I_1} + 2\alpha_{\phi}^2\cos^2\fibredirection\pdiff{W}{I_4} \right]\,. \label{Cpstphi}
  \end{equation}
\end{subequations}
This is a consequence of standard shell theory assumptions concerning the
shell's thin aspect ratio \cite{Green1970}, similar to the approach used,
without fibre-reinforcement, in \cite{Tongen2006} and discussed with
fibre-reinforcement in \cite{Holzapfel2010}. Further details are provided in
\cref{app:stresses}.

Several different strain-energy densities have been used to model sclera
\cite{Coudrillier2013,Grytz2014,copije15}. Keeping with our minimal approach,
here we adopt the simplest strain-energy density for an incompressible elastic
fibre-reinforced material. Following many authors, we use a reinforced
neo-Hookean strain-energy density of the form
\begin{equation}
 W = C(I_1-3) + \frac{D}{4}\left(\maxzero{I_4-1}\right)^2\,, \label{CPsi}
\end{equation}
where the first term represents the isotropic contribution to the stress and
the second term describes the fibre reinforcement. Both $C$ and $D$ are
material parameters with units of stress, and $C$ can be identified with half
of the shear modulus in an isotropic neo-Hookean material. We take $C$ within
the range of values reported in \cite{Girard2009b}, whilst a range of values
for $D$ will be considered. In future work, both $C$ and $D$ could be allowed
to vary with space and time, modeling for instance scleral hardening or
softening with age or emmetropization, respectively. If the strains remain
sufficiently low, this strain energy represents the dominant contribution of
more sophisticated strain-energy functions that only depend on $I_{1}$ and
$I_{4}$, as it can be obtained as the first two terms in a systematic
expansion of a potential $W=W(I_{1},I_{4})$. We also note that the presence of
the invariant $I_{2}$ is known to be important in shear problems, though is
absent from this system due to axisymmetry.

Substituting for $W$ and noting that incompressibility ensures that
$\alpha_s\alpha_{\phi}\alpha_n=1$, we find
\begin{subequations}
\label{Ct}
\begin{equation}
  t_s = 2H\left(C\left(\frac{\alpha_s}{\alpha_{\phi}}-\frac{1}{(\alpha_s\alpha_{\phi})^3}\right) + D[I_4-1]^+\frac{\alpha_s}{\alpha_{\phi}}\sin^2\fibredirection \right)\,,\label{Cts}
\end{equation}
\begin{equation}
  t_{\phi} = 2H\left(C\left(\frac{\alpha_{\phi}}{\alpha_s}-\frac{1}{(\alpha_s\alpha_{\phi})^3}\right)+ D[I_4-1]^+\frac{\alpha_{\phi}}{\alpha_s}\cos^2\fibredirection \right)\,.\label{Ctphi}
\end{equation}
\end{subequations}

\subsection{Growth rate}
A plethora of stimuli and responses have been proposed that could combine in
the human sclera to control emmetropization. Our model allows us to easily
simulate and compare these hypotheses, but in its first exposition we restrict
our attention to a simple scenario in order to illustrate the capabilities of
the framework. Here, the local growth rate, $\eta$, represents the rate at
which a region of the sclera will grow in order to relieve the local stretch.
We decompose this growth rate into the product of an intrinsic capacity for
growth, $g_c$, and a stimulus response, $g_v$. Explicitly, we write
\begin{equation}
 \eta(\sigma,t) = g_c(\Sigma)g_v(\sigma,t)\,, \label{etadef}
\end{equation}
where $g_c$ is a material property dependent only on
$\Sigma=\Sigma(\sigma,t)$. As we are not modeling the growth of the cornea,
matching at the front of the eye requires $g_c\to0$ as $\Sigma\to L$, leading
us to pose the phenomenological functional form
\begin{equation}
  g_c(\Sigma) = \frac{\eta_0}{2}\left(1 + \tanh\left(\frac{\growingzone-\Sigma}{\delta}\right) \right)\,,\label{gc}
\end{equation}
where the location and extent of the growing zone are governed by
$\growingzone$ and the decay away from this zone as we move towards the
anterior sclera is given by $\delta$. The parameter $\eta_0$ has units of
inverse time and quantifies the maximum growth rate of the sclera. The
stimulus dependence of the growth rate, $g_v$, is defined in terms of the
discrepancy between the current, deformed position of the sclera and some
target surface, which will be defined by the optical properties of the eye in
its configuration at time $t$. Thus, $g_v$ is better understood in the form
\begin{equation}
 g_v(\sigma,t) = g_v(r(\sigma,t),z(\sigma,t),t),
\end{equation}
recalling that $s=s(\sigma,t)$. 

Adopting the hypothesised wavelength-dependent growth response of the sclera,
we suppose that the retina can detect the blurring of red and blue light that
reaches it, and that this stimulates or inhibits a growth response. We model
this simply, computing the best-focus surface for blue light as described in
\cref{app:optics}, then defining the growth response based on the location of
the retina relative to this target surface. We thus have two cases to
distinguish. If a material point on the retina is in front of the best-focus
surface for blue light, experiencing hyperopic defocus, then growth is
triggered and its amplitude $g_v$ is set to be the distance between the
material point and the closest point on the blue surface. Alternatively, if a
point on the retina is behind the best-focus surface for blue light,
experiencing myopic defocus, then growth is stopped and we set $g_v$ to zero.
After simulation, the position of the retina is compared with the best-focus
surface for red light to determine if the eye is emmetropic. In particular, if
the retina lies between the best-focus surfaces then we will term the eye
emmetropic, whilst we will describe it as myopic if the retina is behind the
best-focus surface for red light and hyperopic if the retina lies in front of
the best-focus surface for blue light.

\subsection{Model reduction and summary}
\label{sec:reduction}
First, we note that $q_s$ appears in the model only in combination with $r$,
so we replace these terms with the new variable $Q=rq_s$. It is also
convenient to write the equations of mechanical equilibrium in terms of the
grown, undeformed arclength $\sigma$, so that, making use of \cref{G3a},
\cref{G1,G2a,Mfull} become
\begin{subequations}
\label{MG}
\begin{align}
 \pdiff{r}{\sigma} &= \alpha_s\cos\theta\,, \label{MGa} \\
 \pdiff{z}{\sigma} &= -\alpha_s\sin\theta\,, \label{MGb} \\
 \pdiff{\theta}{\sigma} &= \alpha_s\kappa_s\,, \label{MGc} \\
 \pdiff{Q}{\sigma} &= \alpha_s\left[ r\iop - \kappa_srt_s - \kappa_{\phi}rt_{\phi} \right]\,,\label{MQ}\\
 \pdiff{t_s}{\sigma} &= \frac{\alpha_s}{r}\left[ \kappa_sQ + \left(t_{\phi}-t_s\right)\cos\theta \right],,\label{Mts}\\
 \pdiff{m_s}{\sigma} &= \frac{\alpha_s}{r}\left[ (m_{\phi}-m_s)\cos\theta - Q \right]\,.\label{Mm}
\end{align}
\end{subequations}
Substituting our constitutive assumptions for the bending moments into
\cref{Mm}, we find
\begin{equation}
 \pdiff{\kappa_s}{\sigma}= \alpha_s\left(\frac{(\kappa_{\phi}-\kappa_s)}{r}\cos\theta - \frac{Q}{rE_b} \right).\label{Sk}
\end{equation}
\Cref{MGa,MGc,MQ,Mts,Sk} form a system of five ordinary differential equations
in the five unknowns $r$, $\theta$, $\alpha_s$, $Q$ and $\kappa_s$, each as a
function of $\sigma$, noting that $t_s$ may be written as a function of
$\alpha_s$. Following their solution, the variables $z$ and $s$ can be
calculated using \cref{G3a} and \cref{MGb}, i.e.~ $z$ and $s$ decouple. In the
absence of experimental data to guide a choice of growth rate, $\eta_0$, we
nondimensionalise time by a typical timescale of growth, on the order of
years, hereafter considering both time and growth rate to be dimensionless.
Accordingly, we present simulations over a unit time interval.

\subsection{Initial and boundary conditions}
\label{sec:ICBC}
The ungrown unloaded reference configuration is parameterised by the arclength
$\Sigma\in[0,L]$, where $\Sigma=L$ identifies the location where the sclera
meets the cornea. The growing reference domain is thus
$\sigma\in[0,\sigma(L,t)]$. For simplicity, we assume that the original
unloaded configuration has a spherical shape of radius $R_0$, so that
$R=R_0\sin(\Sigma/R_0)$. Thus, the initial conditions for \cref{Grp,Grs} are
simply given by
\begin{subequations}
\begin{align}
 \rho(\Sigma,0) & = R_0\sin\left(\frac{\Sigma}{R_0}\right)\,, \\
 \sigma(\Sigma,0) & = \Sigma\,.
\end{align}
\end{subequations}
We must also provide the initial scleral thickness and fibre orientation. In
our simulations, we consider both shells of uniform thickness and others with
a more biologically realistic profile, with the sclera being thickest at the
anterior pole, thinning towards the equator, then thickening towards the
limbus. We give the explicit form of this non-uniform $H$ in \cref{app:ICBC},
in line with \cite{Fatt1992}. Based on observations in
\cite{Girard2011,Grytz2014}, we consider a fibre orientation that has greater
reinforcement in the $\vec{e}_s$ direction around the equatorial region of the
sclera, but where fibres are predominantly aligned in the $\vec{e}_{\phi}$
direction at the limbus and near the posterior pole, again given explicitly in
\cref{app:ICBC}.

At the back of the eye, where $\sigma=0$, we impose
\begin{subequations}
\label{backBC}
\begin{align}
  r(0,t) & = 0\,, \\
  \theta(0,t) & = 0\,, \\
  Q(0,t) & = 0\,,
\end{align}
\end{subequations}
ensuring continuity of the shell and its slope, with the condition on $Q$
being a consequence of requiring the normal shear stress $q_s$ to be bounded
as $\sigma\rightarrow0$.

At $\sigma = \sigma(L,t)$, we require the sclera to match smoothly to the
cornea. The cornea is a deformable material and therefore should stretch when
the intraocular pressure $\iop$ is raised, suggesting that we should have some
pressure-dependent boundary condition. Instead of explicitly modeling the
anterior eye, we model the cornea as a spherical material in its reference
configuration. When pressure is increased, this surface deforms as detailed in
\cref{app:ICBC}, simply expanding as a spherical shell. The end points of the
cornea are then used as boundary conditions for the deforming sclera, which,
as $\iop$ is held constant, are constant in time. By matching the surfaces, we
have
\begin{subequations}
\label{frontBC}
\begin{align}
 r(\sigma(L,t),t)&=r^{\star}, \\
 \theta(\sigma(L,t),t)&=\theta^{\star},
\end{align}
\end{subequations}
where the dependence of $r^{\star}$ and $\theta^{\star}$ on the system
parameters is explained in \cref{app:ICBC}. After solving
\cref{MGa,MGc,MQ,Mts,Sk} subject to the five boundary conditions in
\cref{backBC} and \cref{frontBC}, we integrate \cref{G3a} and \cref{MGb} to
obtain $s$ and $z$. We set the centreline of the deformed corneal shell to be
the origin of the frame, so that
\begin{subequations}
\begin{align}
& z(\sigma(L,t),t)=z^{\star},\\
& s(0,t)=0,
\end{align}
\end{subequations}
where $z^{\star}$ is defined in \cref{app:ICBC}. Details of the implementation
are given in \cref{app:technical}, with typical parameter values reported in
\cref{tab:param}.

\section{Numerical explorations}
\label{sec:numerics}
\subsection{Reproducing ocular geometry}
\begin{figure}[t]
\centering
\begin{overpic}[width=0.22\textwidth]{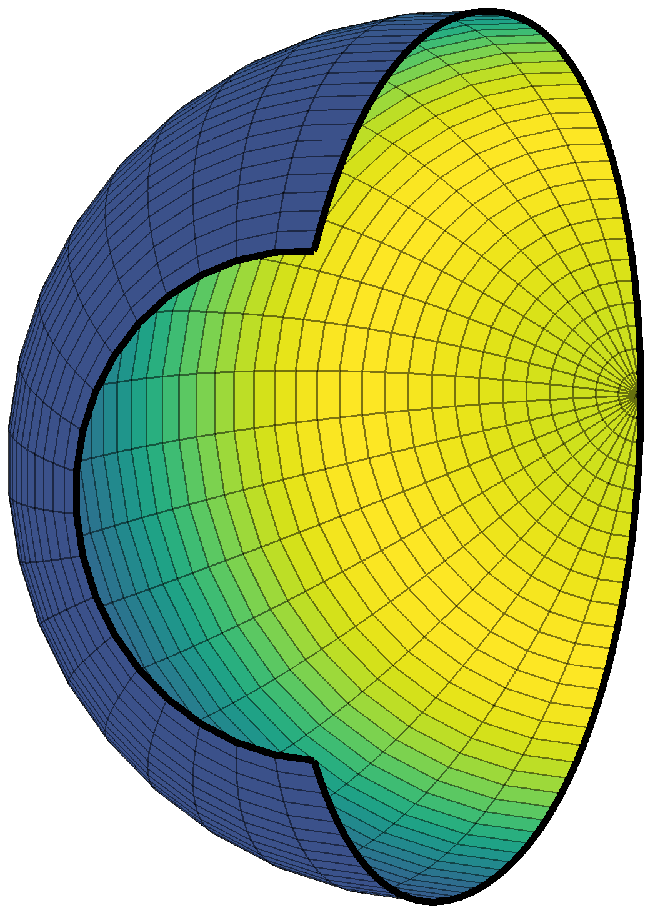}
 \put(0,90){\footnotesize $a)$}
 \put(40,-2){\footnotesize $t=0.1$}
\end{overpic}
\begin{overpic}[width=0.22\textwidth]{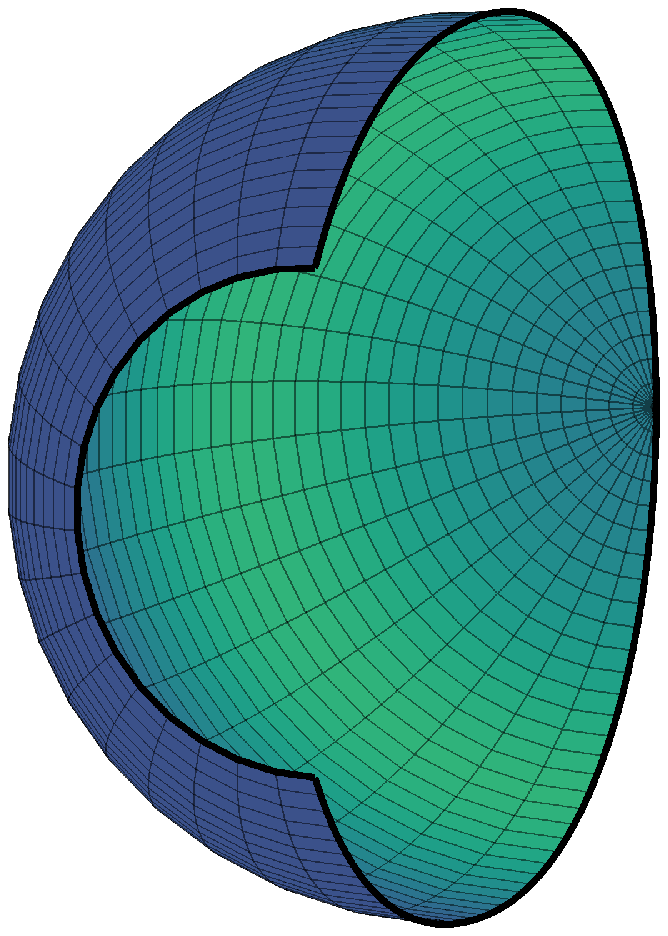}
 \put(0,90){\footnotesize $b)$}
 \put(40,-2){\footnotesize $t=0.2$}
\end{overpic}
\begin{overpic}[width=0.22\textwidth]{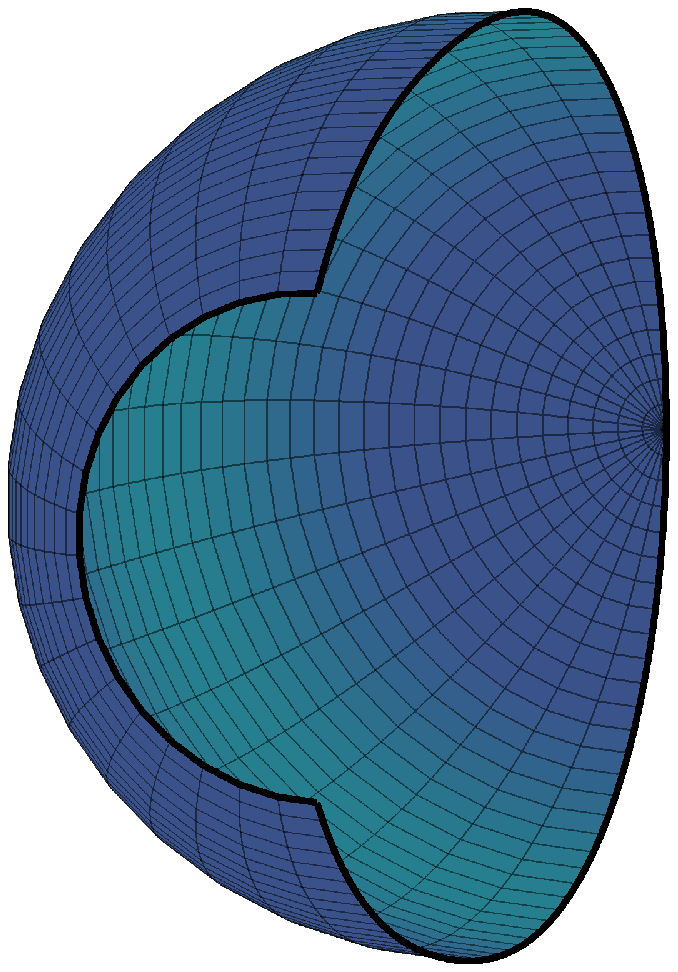}
 \put(0,90){\footnotesize $c)$}
 \put(40,-2){\footnotesize $t=0.4$}
\end{overpic}
\begin{overpic}[width=0.22\textwidth]{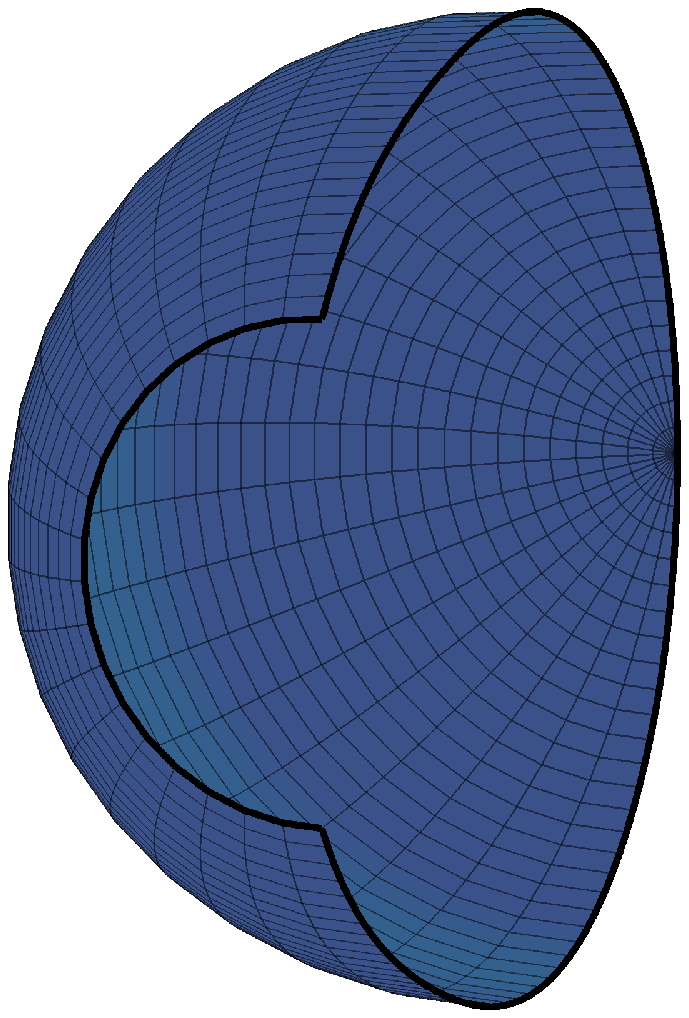}
 \put(0,90){\footnotesize $d)$}
 \put(40,-2){\footnotesize $t=1.0$}
\end{overpic}
\includegraphics[width=0.06\textwidth]{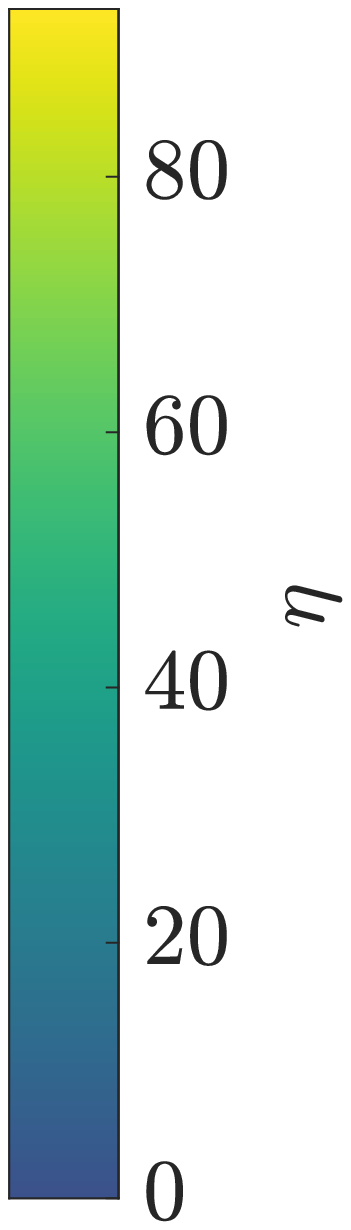}
\\
\vspace*{0.5cm}
\begin{overpic}[width=0.99\textwidth]{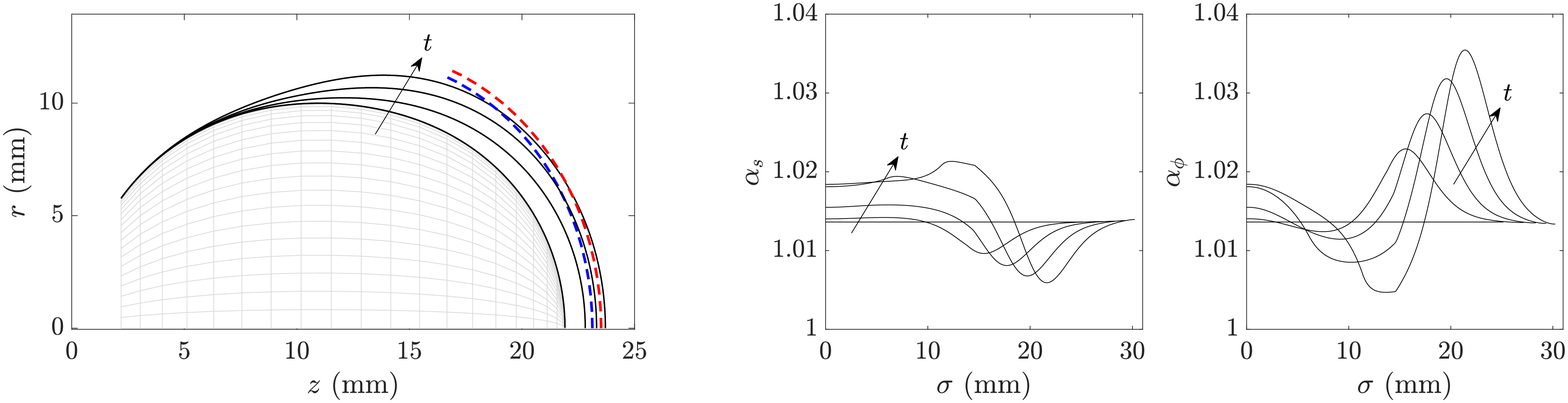}
\put(1,25){e)}
\put(47,25){f)}
\put(74,25){g)}
\end{overpic}
\caption{Numerical solutions of \cref{MGa,MGc,MQ,Mts,Sk}, showing growth of a
uniform-thickness, fibre-free sclera. (a-d) The scleral shell throughout the
growth and deformation, shown cut at select timepoints, shaded by the growth
rate $\eta$. (e) The evolution of the retina over time, corresponding to the
sclera shown in (a-d), with the target best-focus surface for blue light and
the best-focus surface for red light also shown. Snapshots of the stretch in
the (f) $\vec{e}_s$ direction and (g) $\vec{e}_{\phi}$ direction at times
$t=0,0.1,0.2,0.4,1$, with arrows indicating increasing time, noting the
uniform initial stretches due to the initially uniform shell thickness. The
regions of largest growth rate migrate away from the posterior eye over time,
with the posterior region rapidly nearing the target surface and thus
experiencing slower growth. Despite this negative feedback, the axial length
of the eye increases past the best-focus surface for red light near the fovea,
resulting here in myopia. Parameter values used in this simulation are
$C=100$kPa, $D=0$, $H=0.65$mm, $E_b=18$kPamm$^3$, $\eta_0=70$mm$^{-1}$,
$\growingzone=8$mm and $\delta=4$mm.}
\label{fig:const}
\end{figure}

In \cref{fig:const} we present the simulated growth and deformation of a
sclera with uniform reference thickness and no fibre reinforcement. In
\cref{fig:const}a-d we show half of the scleral shell, shaded by the growth
rate $\eta$, with the corresponding evolution of the retina relative to the
best-focus surfaces for blue and red light shown in \cref{fig:const}e. What is
most notable is the qualitatively plausible shape attained by the sclera and
retina, with the local optically driven growth law apparently sufficient to
produce realistic morphologies via the presented minimal model. We also
observe that the region of fastest growth migrates away from the posterior
sclera, in line with the continued progress of the posterior retina towards
the best-focus surface for blue light. Indeed, at later times the axial growth
of the eye is being driven by growth in regions away from the posterior
sclera, concentrated in a band nearing the scleral equator, with the rearmost
portion of the retina having moved behind the best-focus surface for blue
light and therefore inhibiting local growth. In this case, this leads to a
myopic eye, with the posterior retina at $t=1$ having moved past the
best-focus surface for red light.

Figures \ref{fig:const}f and \ref{fig:const}g depict the elastic stretches in
the $\vec{e}_s$ and $\vec{e}_{\phi}$ directions, initially uniform and equal
due to the homogeneous spherical initial condition. The non-uniform growth of
the sclera breaks this homogeneity, resulting in stretches that vary around
the eye. Perhaps unsurprisingly, we note that regions of high $\alpha_s$ are
approximately coincident with regions of reduced $\alpha_{\phi}$, and vice
versa.

\begin{figure}[t]
\centering
\begin{overpic}[width=0.99\textwidth]{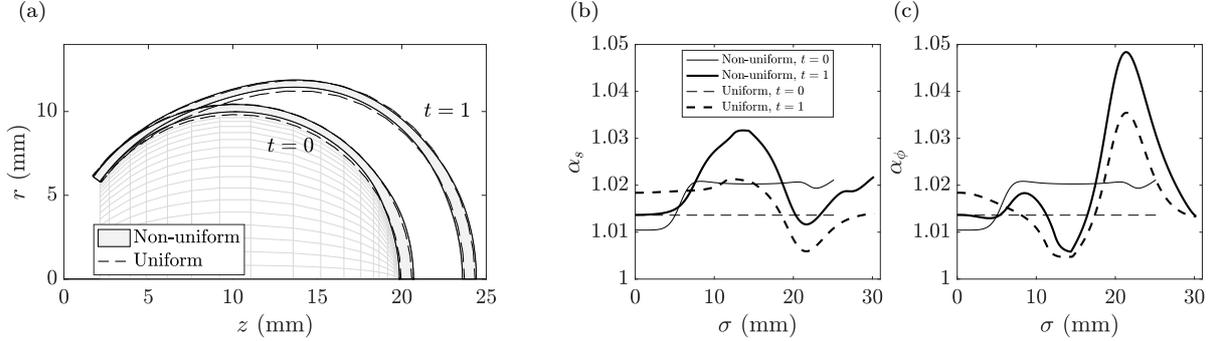}
\put(1,27){(a)}
\put(47,27){(b)}
\put(74,27){(c)}
\end{overpic}
\caption{Growth of a fibre-free sclera with spatially varying unloaded
thickness, with $H$ as defined in \cref{Hdefn}, compared to a sclera with
uniform thickness. (a) The initial and final deformed sclera, where the solid
and dashed lines are the surfaces of the sclera in the non-uniform and uniform
cases, respectively, with the non-uniform case shaded. The stretches in the
(b) $\vec{e}_s$ and (c) $\vec{e}_{\phi}$ directions are shown as solid and
dashed curves for the non-uniform and uniform sclera, respectively, with
lighter curves corresponding to the initial condition and heavier curves
corresponding to the final configuration. The difference in scleral reference
structure appears to minimally impact the evolving shape of the eye, though
differences in the associated stretches are evident. In particular, the
non-uniform reference configuration serves to reduce both of these stretches
near the posterior eye ($\sigma=0$). The parameters used in this simulation
are $\iop=2$kPa, $C=100$kPa, $D=0$, $E_b=18$kPamm$^3$, $\eta_0=70$mm$^{-1}$,
$\growingzone=8$mm and $\delta=4$mm.}
\label{fig:thick}
\end{figure}

\subsection{Effects of a non-uniform reference thickness}
Now incorporating the non-uniform reference thickness of the sclera, as given
explicitly in \cref{Hdefn}, we evaluate the differences in the morphology and
elastic stretches between this and the previous uniform case of
\cref{fig:const}. The initial and final loaded configurations of the sclera
are shown in \cref{fig:thick}a, where the solid black lines denote the upper
and lower surfaces of the sclera, whilst the dashed lines correspond to the
case with uniform reference thickness. We observe only a small difference in
scleral and retinal positioning due to the non-uniform reference thickness,
though the retina is marginally shifted forwards compared to the uniform case
due to the increased scleral thickness in the posterior eye. However, the
elastic stretches associated with these final deformed configurations, shown
in \cref{fig:thick}b and \cref{fig:thick}c, are more significantly altered,
with the maximum stretches in the posterior sclera reduced in the variable
reference thickness case. The variation in the stretches can be attributed
directly to the non-uniform reference thickness, with thinner regions
experiencing comparatively larger stretches.

\subsection{Subtleties of fibre reinforcement}
\begin{figure}[t]
\centering
\begin{overpic}[width=0.8\textwidth]{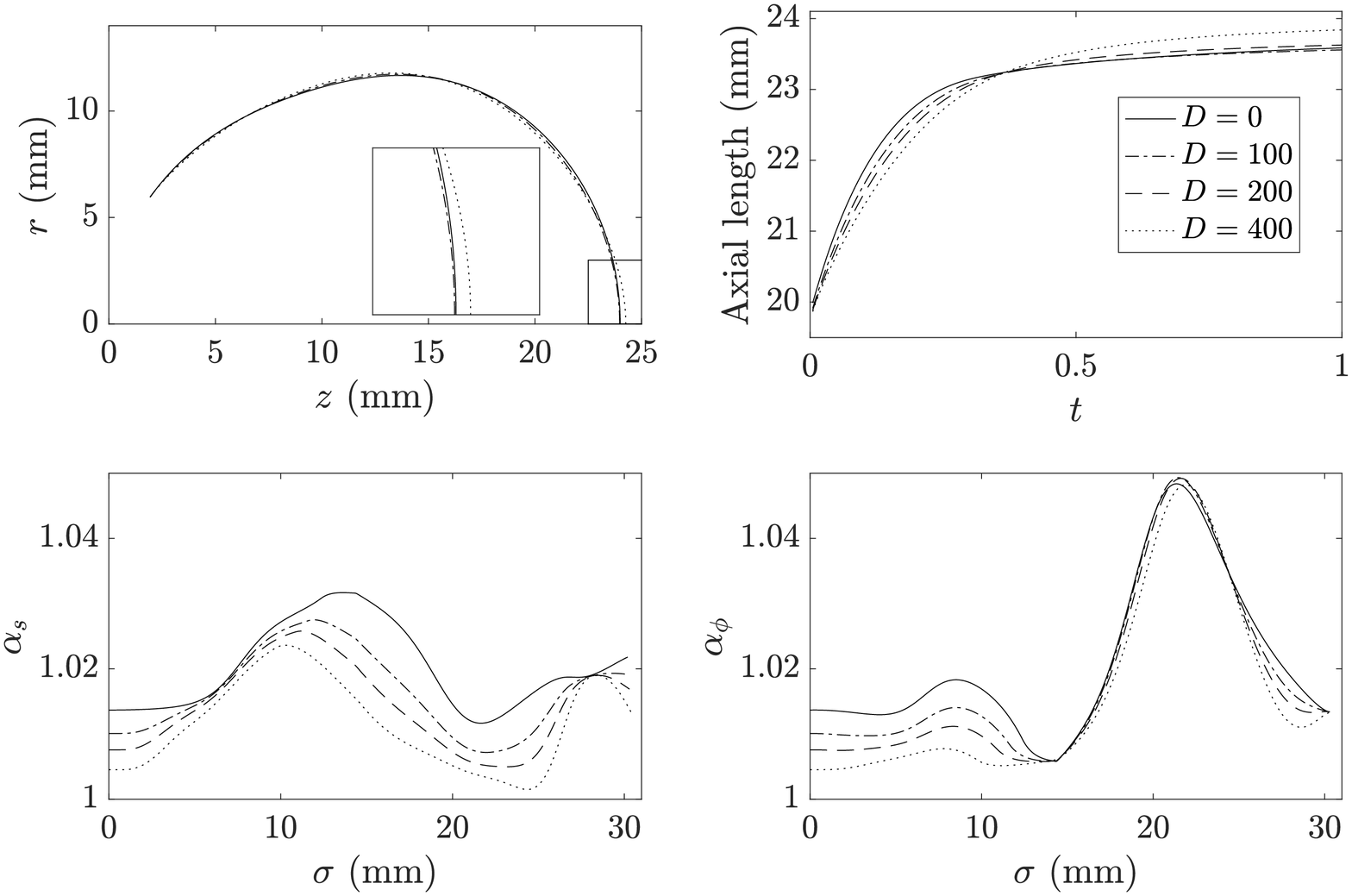}
\put(5,67){(a)}
\put(55,67){(b)}
\put(5,33){(c)}
\put(55,33){(d)}
\end{overpic}
\caption{Effects of fibre-reinforcement on scleral growth. (a) The final
deformed position of the sclera for various levels of fibre reinforcement.
Inset is a magnified view of the posterior sclera, which reveals that the
strongest fibres result in an increased axial length and reduced scleral angle
at the back of the eye, though both changes are marginal. (b) The evolution of
the axial length of the eye, measured from $z=0$ to the surface of the
posterior retina, revealing slower and prolonged growth for the stronger
fibres, though this effect is non-monotonic in fibre strength (cf $D=0$ and
$D=100$). The stretches in the (c) $\vec{e}_s$ and (d) $\vec{e}_{\phi}$
directions typically decrease with increased fibre strength, though a complex
relationship appears around $\sigma=21$, where the stretch in the
$\vec{e}_{\phi}$ direction increases with strong fibre reinforcement. The
scleral thickness is given by \cref{Hdefn} and the fibre orientation is given
by \cref{adefn}. The parameters used are $\iop=\SI{2}{\kilo\pascal}$,
$C=\SI{100}{\kilo\pascal}$, $E_b=\SI{18}{\kilo\pascal\milli\metre\cubed}$,
$\eta_0=\SI{70}{\per\milli\metre}$, $\growingzone=\SI{8}{\milli\metre}$,
$\delta=\SI{4}{\milli\metre}$, taking the non-uniform reference thickness $H$
of \cref{Hdefn}. The fibre strength ranges from
$D=$\SIrange{0}{400}{\kilo\pascal}, as stated in the legend without units,
though the $D=\SI{200}{\kilo\pascal}$ simulation has been omitted from (a) and
(b) for clarity.}
\label{fig:fibre}
\end{figure}

Retaining the non-uniform scleral reference thickness considered above, we
incorporate the fibre orientation prescribed in \cref{adefn} with a range of
fibre strengths $D$, presenting a selection of grown deformed sclera in
\cref{fig:fibre}a. Immediately evident is the minimal effect that this fibre
reinforcement has on the final scleral morphology, with only small differences
present. That being said, the highest degree of reinforcement considered here
results in increased axial length and a reduced scleral angle at the
posterior, as shown inset in \cref{fig:fibre}a. The axial growth dynamics are
illustrated in \cref{fig:fibre}b, from which we again see the increased length
of the most reinforced sclera, though resultant from a lower rate of growth
over an extended period of time compared to sclera with lower levels of
reinforcement. Further, whilst the most reinforced shell has the greatest
final length, the shell with only slight reinforcement exhibits a reduced
length compared to the fibre-free shell. Thus, there is a non-monotonic
response of the scleral morphology to the strength of fibre reinforcement. The
existence of such a complex response of a scleral shell to fibre reinforcement
is further illustrated in \cref{fig:fibre}c and \cref{fig:fibre}d, with the
stretch in the $\vec{e}_{\phi}$ direction in the anterior sclera not following
the overall trend of being reduced by increased reinforcement.

\begin{figure}[t]
\centering
\begin{overpic}[width=0.9\textwidth,permil]{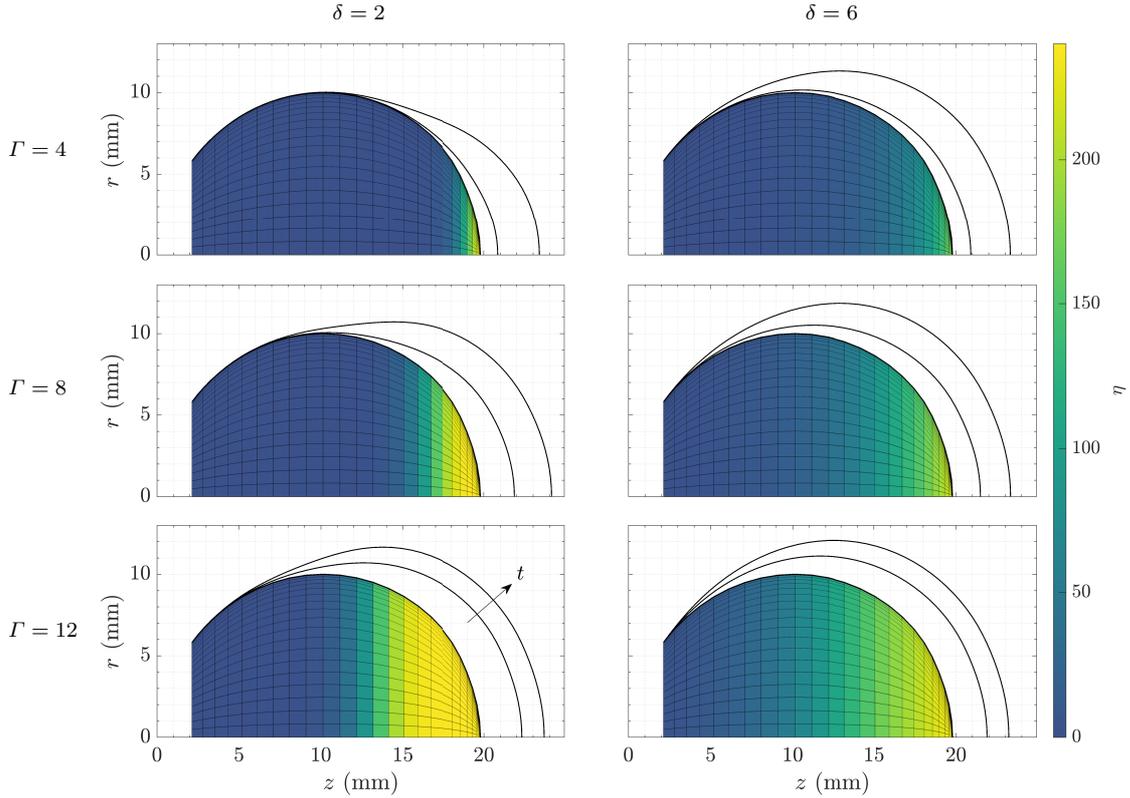}
\put(-40,624){$\growingzone=4$}
\put(-40,403){$\growingzone=8$}
\put(-40,179){$\growingzone=12$}
\put(258,750){$\delta=2$}
\put(692,750){$\delta=6$}
\end{overpic}
\caption{Effect of the growth zone on the dynamics and position of the retina,
shown in the initial configuration, at $t=0.1$ and at $t=1$, shaded by growth
rate $\eta$. Varying both the location of the growing region, $\growingzone$,
and the rate of transition between low and high growth, $\delta$, we observe a
range of growth dynamics and final configurations of the retina. A smoother
transition between regions of fast and slow growth appears to result in a more
spherical retina, whilst increasing $\growingzone$ is associated with rapid
initial growth rate, which is discernible from the location of the $t=0.1$
curves relative to the shared initial configuration. Despite the initially
increased growth rate, higher values of $\growingzone$ do not always result in
increased axial length, as more clearly displayed in \cref{fig:axial}. Here,
we have taken the non-uniform reference scleral thickness of \cref{Hdefn},
considering fibre reinforcement as described in \cref{adefn}. The material
parameters used here are $\iop=2$kPa, $C=100$kPa, $D=100$kPa and
$E_b=18$kPamm$^3$, with $\eta_0=\SI{70}{\per\milli\metre}$. Both
$\growingzone$ and $\delta$ have units of millimetres.}
\label{fig:grow}
\end{figure}

\subsection{Impacts of growth laws}
\begin{figure}[t]
\centering
\vspace{0.5mm}
\begin{overpic}[width=0.8\textwidth,permil]{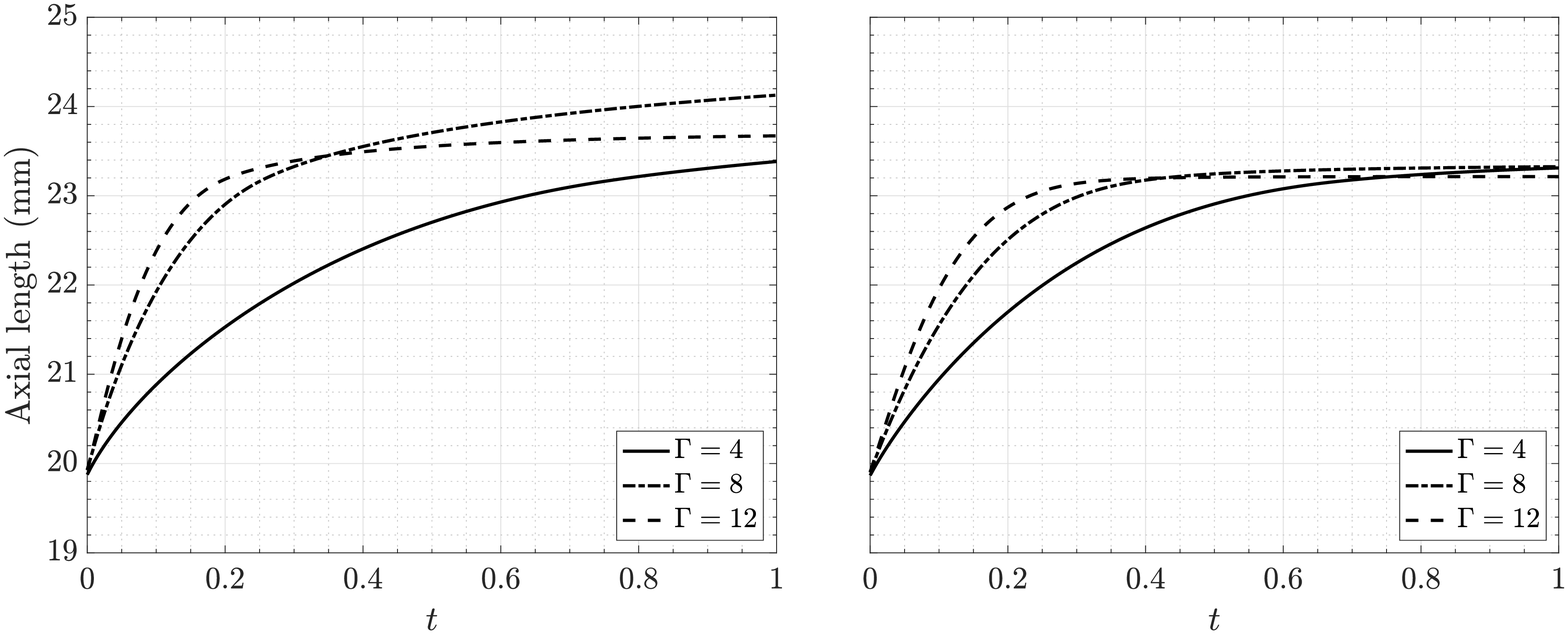}
\put(246,408){$\delta=2$}
\put(747,408){$\delta=6$}
\end{overpic}
\caption{Change in axial length of the eye over time for various regions of
growth, corresponding to the configurations shown in \cref{fig:grow}. With the
intrinsic capacity for growth specified as in \cref{gc}, we see that increased
$\growingzone$ results in a faster initial rate of growth, though for high
$\delta$ the maximum axial length is approximately independent of the location
of maximal intrinsic growth, $\growingzone$. Indeed, the overall axial length
at $t=1$ appears to be robust to variations in the parameters, with all but
the $(\growingzone,\delta)=(12,2)$ case being within a range of
\SI{3}{\milli\metre}. The material parameters used are as described in
\cref{fig:grow}. Both $\growingzone$ and $\delta$ have units of millimetres.}
\label{fig:axial}
\end{figure}

In order to investigate the role of the intrinsic growth capacity, we consider
a variety of parameter combinations $(\growingzone,\delta)$, which
parameterise the location of the growing region and the smoothness of the
transition between regions of high and low growth capacity, respectively.
\Cref{fig:grow} shows significant changes in the morphology of the grown
deformed retina as we vary these parameters. At low values of $\delta$, when
the transition from low to high growth capacity is rapid, we observe a marked
change in retinal shape as the location of the growing zone is moved away from
the posterior eye as $\growingzone$ increases, with an upwards bump developing
for $\growingzone=\SI{12}{\milli\metre}$. As $\growingzone$ is increased, we
also note a change in the rate of growth, with the eye rapidly growing for
high values of $\growingzone$, though still attaining similar axial lengths at
$t=1$. This is shown explicitly in \cref{fig:axial}, in which the axial growth
rate can be seen to be strongly dependent on $\growingzone$ when
$\delta=\SI{2}{\milli\metre}$. Here, growth at the peripheral retina drives
the axial progression at late stages of development, as seen earlier in
\cref{fig:const}. Here, we note another non-monotonic dependence of the ocular
morphology on the parameters, that of final axial length on the location of
the growing region.

Surprisingly, similar dependence on $\growingzone$ is not present when
considering $\delta=\SI{6}{\milli\metre}$, with the axial length of the eye at
$t=1$ approximately independent of $\growingzone$ now that the intrinsic
growth capacity transitions more gradually. Returning to \cref{fig:grow}, we
also see that the final shape of the retina is largely unaffected by changes
to $\growingzone$, despite the change in growth rate. Thus, the smoothness of
the transition between regions of high and low growth capacity appears
dominant over the location of maximal growth and the associated rate of
development. This suggests partial robustness of the retinal/scleral
development process to the details of growth, further supported by the
approximately consistent axial length seen throughout each of these
simulations, with the exception of
$(\growingzone,\delta)=(\SI{8}{\milli\metre},\SI{2}{\milli\metre})$.

Additionally, the retinal morphology takes on an approximately spherical form
when $\delta=\SI{6}{\milli\metre}$, in stark contrast to the varied shapes
seen for $\delta=\SI{2}{\milli\metre}$. This suggests that the smoother
transition of the growing region in the former seemingly drives this more
uniform growth and deformation, consistent with intuition and distinct from
the sharply varying growth dynamics present when
$\delta=\SI{2}{\milli\metre}$.

\section{Discussion}
\label{sec:discussion}
This work has described and showcased an idealised model for the growth and
development of the primary structural component of the eye, the sclera. Under
the assumption of morphoelasticity and axisymmetry, we have derived a simple
yet detailed model in which the growth mechanics of the thin scleral shell may
be readily coupled to external stimuli or material properties. Reducing to a
simple system of five quasistatic ordinary differential equations, with the
growth and elastic timescales separated by orders of magnitude, this flexible
framework may be solved with standard numerical methods, achievable without
significant computational cost or optimisation. This model is therefore
well-suited to explorative studies of ocular development, sacrificing the
accuracy of geometrically refined models in favour of rapidly querying the
fundamental principles that link the growth and deformation during
emmetropization.

In order to showcase the flexibility and utility of this approach, we have
included and briefly explored the effects of multiple mechanisms and features
of the developing eye, focussing in particular on a hypothesised driver of
growth. With the optical properties of the mid-growth eye either stimulating
or inhibiting the growth of the model sclera based on the detection of
hyperopic blur, we have seen that this hypothesis can lead to qualitatively
realistic ocular morphologies for typical and estimated parameter values.
Having sought throughout to impose the simplest plausible assumptions and
constitutive laws for the geometry, mechanical properties and growth kinetics
of the sclera, we have therefore seen that these are sufficient to
phenomenologically capture the growth dynamics of the eye. In particular, the
prescribed local growth law was sufficient to achieve an appropriate global
response to external stimuli, forming eyes of a plausible size and shape for
focused vision.

Further, the observation of qualitative realism was generally found to be
robust to changes in the details of the growth specification, though
significant variation within this broad class was observed when modifying the
intrinsic growth capacity of the sclera. This resulted in a range of
configurations at maturity, most notable being the approximately spherical
shapes observed when local growth varied more slowly over the sclera.
Surprisingly, the inclusion of fibres had little effect in comparison to that
of varying the growth specification. Indeed, the key property of the mature
eye, the axial length, was largely unchanged by varying the degree of
reinforcement. However, the small observed changes exhibited a non-monotonic
response to increases in reinforcement strength, suggesting the existence of a
complex relationship between shell structure and growth dynamics that future
study is expected to explore in detail.

Our presented simulations posit a finite time interval in which the sclera can
grow; thereafter, all growth stops. In practice, the growth rate may vary more
smoothly with age and there may be stages in postnatal developmental when the
eye experiences more rapid growth, for example. These non-uniform growth
periods could be considered via a simple extension of the framework developed
in this work, and therefore could easily be investigated. Additional readily
realisable refinements could include coupling this model with the evolving
optical properties of the front of the eye, with an interesting application
being the evolution of best-focus surfaces during childhood and how the sclera
grows in response to these changes. Another focus that merits further
theoretical investigation concerns the mechanism by which the sclera detects
and responds to visual information, for example the effects of individual
variation in photoreceptor topography and its link to the axial length of the
eye \cite{Wang2019}. If chromatic effects are demonstrated to be significant
in emmetropization, one can envisage the design of corrective eyewear, guided
by mathematical modeling, that specifically tailors the shape of image
surfaces to moderate the growth of the eye.

In summary, we have presented a simple morphoelastic model of the complex
multifaceted process of emmetropization. The model provides a step towards
improving understanding of this developmental process, demonstrating that
local growth laws can lead to qualitatively realistic morphologies of the
elastically deformed eye, whilst enabling simple yet detailed future
explorations of varied hypotheses for ocular development.

\begin{acknowledgements}
The research leading to these results has received funding from the European
Union Seventh Framework Programme (FP7/2007-2013) under grant agreement no.
309962 (HydroZONES). BJW is supported by the UK Engineering and Physical
Sciences Research Council (EPSRC), Grant No. EP/N509711/1.
\end{acknowledgements}

The computer code used and generated in this work is freely available from
\url{https://gitlab.com/bjwalker/morphoelastic-eye.git}

\appendix
\section{Optical calculations}
\label{app:optics}
\begin{figure}[t]
\centering
\includegraphics[width=0.7\textwidth]{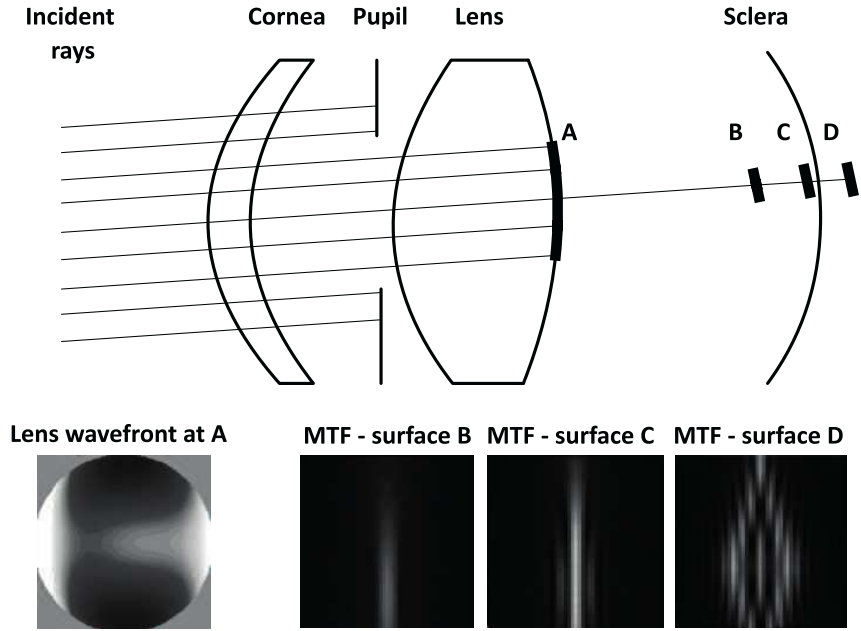}
\caption{Computing the best-focus surfaces. A dense array of parallel,
coherent rays (thin lines) is traced through the anterior optics (cornea,
pupil, lens – medium lines) and the phase of the wavefront emerging through
the posterior surface of the lens is fitted to Zernike polynomial functions.
Bottom – Left: Example of a lens-emerging waveform.  A Kirchhoff integral is
applied to further propagate the wavefront and compute its modulation transfer
function (MTF) on small surfaces perpendicular to the central ray. Bottom –
Right: Example MTFs along probed surfaces. Surface C maximizes the Strehl
ratio and hence corresponds to the best-focus surface for this wavelength.
Computing the best-focus distance for a range of incidence angles (0-40
degrees), assuming axial symmetry, we reconstruct the best-focus surfaces for
two wavelengths: \SI{400}{\nano\metre} (blue) and \SI{600}{\nano\metre}
(red).}
\label{fig:RayTrace}
\end{figure}

The optical components of the anterior eye, the cornea, anterior chamber, lens
and vitreous chamber, focus the light wavefronts that are incident on the eye
on a fictitious curved surface near the retina, which we term the best-focus
surface. The position and shape of this surface are dependent on the
wavelength of the incident light due to chromatic aberrations in the light
focusing components. Modeling the geometrical and optical properties of the
anterior eye as in \cite{Atchison2006}, a raytracing algorithm was employed in
order to compute the individual surfaces of best focus for red and blue
incident light wavefronts, exemplified in \cref{fig:RayTrace}. Dense arrays of
parallel coherent rays were traced through the anterior optics, with the phase
of the wavefront emerging on the posterior surface of the lens fitted to
Zernike polynomial functions. These are propagated via a Kirchhoff integral
and the Strehl ratio is computed on various test surfaces perpendicular to the
central ray. The surface corresponding to the maximum Strehl ratio represents
the best-focus surface, which is constructed for incident angles between 0 and
40 degrees, appealing to assumed axisymmetry. This approach may be readily
extended to include the effects of additional or non-uniform lenses, enabling
the modeling of corrective lenses and their effects on ocular development, for
example.

\section{Initial and boundary conditions}
\label{app:ICBC}
\begin{figure}[t]
\centering 
\begin{overpic}[width=0.7\linewidth]{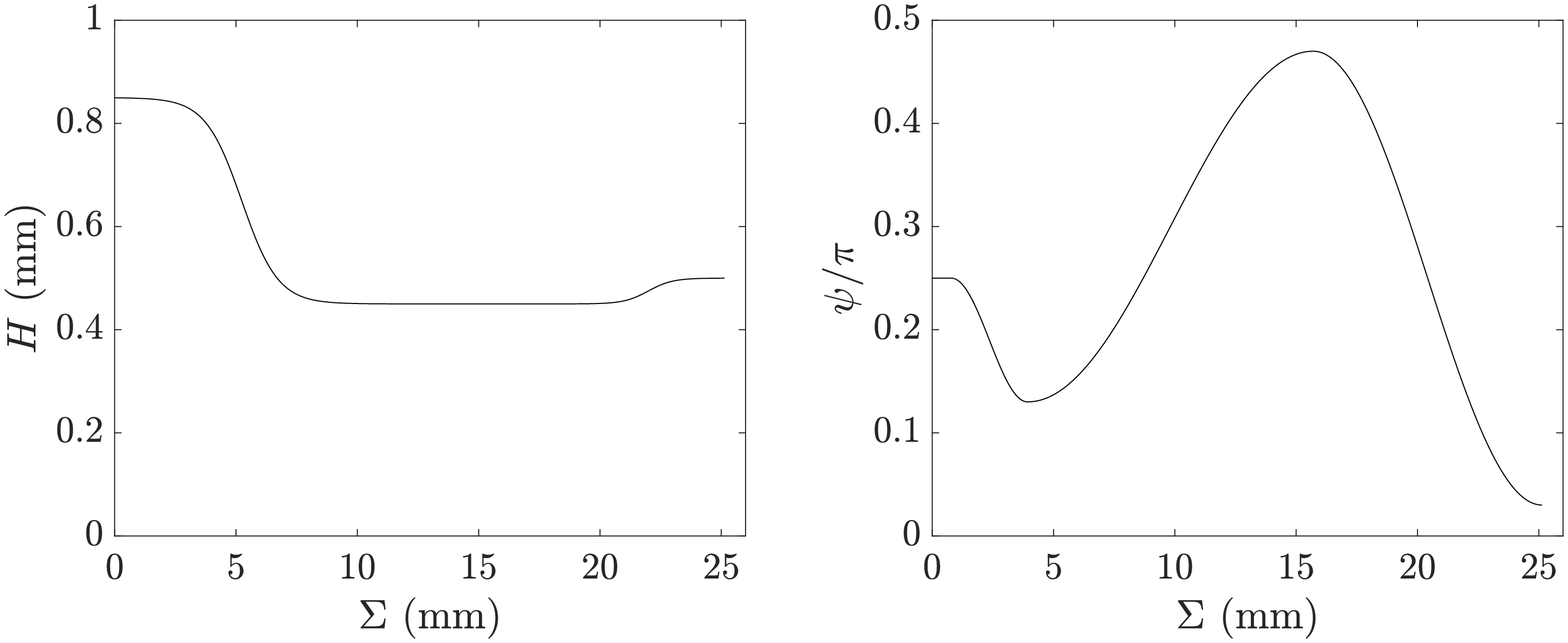}
\put(1,42){(a)}
\put(51,42){(b)}
\end{overpic}
\caption{Reference non-uniform scleral thickness and fibre orientation, shown
in (a) and (b), respectively, from the works of \cite{Fatt1992} and
\cite{Grytz2014}.}
\label{fig:HAdefn}
\end{figure}

When considering a non-uniform scleral thickness, following \cite{Fatt1992} we
prescribe
\begin{equation}
  H(\Sigma) = \left\{\begin{array}{lr}
  0.65-0.2\tanh\left(\frac{\Sigma-0.167\pi R_0}{1.5}\right)\,, &  x\in[0,0.5\pi R_0]\,,\\
  0.475+0.025\tanh(\Sigma-0.7\pi R_0)\,, & x\in(0.5\pi R_0,L]\,,
  \end{array}\right.
  \label{Hdefn}
\end{equation}
as shown in \cref{fig:HAdefn} alongside the initial fibre orientation,
prescribed as
\begin{equation}
  \frac{\psi({\Sigma})}{\pi} = 0.25 + \left\{\begin{array}{lr}
  0\,, & x\in[0,0.025\pi R_0]\,,\\
  0.06\cos\left(\frac{\Sigma-0.025\pi R_0}{0.1R_0} \right)- 0.06\,, & x\in(0.025\pi R_0,0.125\pi R_0]\,,\\
  0.15 - 0.17\cos\left(\frac{\Sigma-0.125\pi R_0}{0.375R_0} \right)\,, & x\in(0.125\pi R_0,0.5\pi R_0]\,,\\
  0.22\cos\left(\frac{ \pi(\Sigma-0.5\pi R_0)}{L - 0.5\pi R_0} \right)\,, & x\in(0.5\pi R_0,L]\,,
  \end{array}\right.
\label{adefn}
\end{equation}
following the observations of \cite{Girard2011,Grytz2014}.

At the anterior point of the sclera, we match the scleral displacement to the
inflation of a thin, spherically symmetric, non-growing shell of uniform
thickness with no fibres, minimally modeling the cornea. Firstly, for this
simple shell, we see that $\alpha_s^c=\alpha_{\phi}^c$, so for notational
convenience we denote the stretch simply by $\alpha$, where the superscript on
all other variables denotes that we are considering the cornea. Since the
corneal reference configuration is spherical, we have
\begin{equation}
 R^c = R_0^c\sin\left(\frac{\Sigma^c}{R_0^c} \right),
\end{equation}
for $\Sigma^c\in[0,2\pi R_0^c]$, where $R_0^c$ is the radius of the sphere. The
position of a point on the inflated sphere is thus given by
\begin{subequations}
\begin{align}
 r^c&=\alpha R_0^c\sin\left(\frac{\Sigma^c}{R_0^c} \right)\,, \\
 z^c&=\alpha R_0^c\cos\left(\frac{\Sigma^c}{R_0^c} \right)+B\,,
\end{align}
\end{subequations}
where $B$ is a constant of integration. The constraint of spherical symmetry
ensures there is no normal shear force, $Q^c=0$, so that the shell deforms as
a membrane. The solution to this problem is presented in \cite{Adkins1952},
where it is shown that
\begin{equation}
 \iop =\frac{4C^cH^c}{\alpha R_0^c}\left(1-\frac{1}{\alpha^6}\right),\label{BCcorn1}
\end{equation}
where $C^c$ is the neo-Hookean constant, $H^c$ is the undeformed thickness and
$R_0^c$ is the undeformed radius of the shell. We take $C^c$, $H^c$ and
$R_0^c$ to have values based on the mechanics of the cornea and we calculate
$\alpha$ for the required pressure difference numerically, restricting
$\alpha\in(1,7^{(1/6)})$ due to the non-injective relation between $\alpha$
and $\iop$. In particular, the upper limit here is the $\alpha$ value
corresponding to the maximum of $1/\alpha-1/\alpha^7$ for $\alpha>1$, it
placing a bound on the maximum pressure difference that we can consider,
though we don't vary the intraocular pressure in this work. Finally,
evaluating the shape of the cornea at the point of attached to be sclera, we
find the boundary conditions for the scleral shell to be
\begin{subequations}
\label{starconstants}
\begin{align}
 r^{\star}&=\alpha R_0^c\sin\left(\frac{L}{R_0^c}\right), \\
 \theta^{\star}&=\frac{L}{R_0^c}, \\
 z^{\star}&=\alpha R_0^c\left(1+\cos\left(\frac{L}{R_0^c}\right)\right).
\end{align}
\end{subequations}

\section{Implementation}
\label{app:technical}
The governing equations presented in \cref{sec:reduction} have a singularity
when $r=0$, so we solve the system numerically on the truncated domain
$\sigma\in[\sigma(\varepsilon,t),\sigma(L,t)]$ for $0<\varepsilon\ll 1$. By
expanding the variables $r$, $\theta$, $\alpha_s$, $\kappa_s$ and $Q$ around
$\sigma=0$ and evaluating at $\sigma=\sigma(\varepsilon,t)$, following
\cite{Woolley2013}, then substituting the expansions into (\ref{MG}{\it
a,c,d,e}) and \cref{Sk} subject to \cref{backBC}, it becomes clear that the
singularity is removable for compatible initial fibre directions. Indeed,
isotropy is required as $\sigma\rightarrow0$ because any preferred direction
is undefined at the pole, and if we do not require
$\fibredirection\rightarrow\pi/4$ as $\sigma\rightarrow0$ then there is a
singularity in the stress as $\sigma\rightarrow0$ in the fibre-reinforced
shells. Intuitively, this is due to the preferred fibre orientation needing to
change direction increasingly quickly as we approach the pole. In order to
circumvent this in all the fibre-reinforced simulations in this work, we
ensure that $\fibredirection\rightarrow\pi/4$ as $\sigma\rightarrow0$, subject
to which the stress is finite and the boundary conditions on $r$ and $\theta$
can be replaced by the notationally cumbersome
\begin{subequations}
\label{BCtrunc}
\begin{align}
 r(\varepsilon)&= \sigma(\varepsilon)\alpha_s(\sigma(\varepsilon)), \\ 
 \theta(\varepsilon)&= \sigma(\varepsilon)\alpha_s(\sigma(\varepsilon))\kappa_s(\sigma(\varepsilon))\,,
\end{align}
\end{subequations}
where we have suppressed the $t$-dependence of all quantities here for
brevity. Now considering $Q$, we further manipulate \cref{MG} to admit the
first integral
\begin{equation}
 Q\cos\theta +  rt_s\sin\theta - \frac{r^2\iop}{2} = A,\label{MFI}
\end{equation}
where $A$ is a constant. Since we require solutions that pass through $r=0$
with $\theta=\pi/2$, we find $A=0$. Thus, evaluating \cref{MFI} at
$\sigma=\sigma(\varepsilon,t)$ provides the analogous truncated boundary
condition for $Q$. Note that whilst it is possible to use \cref{MFI} to
eliminate $Q$ from \cref{MGa,MGc,MQ,Mts,Sk}, preliminary numerical simulations
suggested that it is easier to solve the five ordinary differential equations
than the reduced system. Hence, we retain $Q$ in the governing equations and
use \cref{MFI} as a check on the numerical solutions.

We utilise MATLAB's inbuilt adaptive boundary problem solver \texttt{bvp4c} to
solve equations \cref{MGa,MGc,MQ,Mts,Sk} subject to the boundary conditions
truncated boundary conditions. The initial conditions are provided on a
regular grid for $\Sigma\in[0,L]$ and growth is approximated with an explicit
Euler scheme for \cref{Grp,Grs}. For each simulation, we ensure that the
solution has converged with respect to our choices of grid size, timestep,
truncation point and error tolerances in the solver. For example, the
simulations in \cref{fig:const} were rerun on a refined spatial grid, with a
smaller timestep, with a lower error tolerance in the \texttt{bvp4c} solver,
and with a reduced truncation value $\varepsilon$. The largest relative errors
in the variables $\kappa_s$, $\alpha_s$, $r$ and $\theta$ at $t=1$ in this
refined simulation are $1.7\times10^{-3}$, $2.1\times10^{-5}$,
$9.4\times10^{-4}$ and $7.4\times10^{-4}$ respectively, well below practical
tolerance. Typical parameter values for the simulations in this work are given
in \cref{tab:param}.

\begin{table}
\centering
\begin{tabular}{|l|l|l|}
\hline
Name & Value & Source \\
\hline
$R_0$ & \SI{10}{\milli\metre} & \cite{Gordon1985} \\
$L$ & {$0.8\pi R_0$}\si{\milli\metre} & \cite{Gordon1985} \\
$H$ & \SI{0.65}{\milli\metre} & Based on \cref{Hdefn} \\
$\iop$ & \SI{2}{\kilo\pascal} & \cite{Oyster1999} \\
$C$ & \SI{100}{\kilo\pascal} & \cite{Girard2009b} \\
$D$ & \SIrange{100}{400}{\kilo\pascal} & Estimated \\
$E_b$ & \SI{18}{\kilo\pascal\milli\metre\cubed} & \cite{Howell2009} \\
$\eta_0$ & \SI{70}{\per\milli\metre} & Estimated \\
$\delta$ & \SIrange{4}{12}{\milli\metre} & Estimated \\
$\growingzone$ & \SIrange{2}{6}{\milli\metre} & Estimated \\
\hline
\end{tabular}
\caption{Typical parameter values used in simulations, unless otherwise
specified. $R_0$ and $L$ represent the size of a typical sclera towards the
end of the rapid growth phase of development, at the beginning of the time
interval that we model. The estimate for $C$ is based on values measured in
monkey sclera, albeit with a slightly different constitutive law. The bending
stiffness is estimated from $C$ and $H$ under the assumption of
incompressibility. The corneal parameters $R_0^c$, $H^c$ and $C^c$ are chosen
to match the scleral properties.}
\label{tab:param}
\end{table}

\section{Shell stresses}
\label{app:stresses}
We briefly discuss the standard rationale for the functional form of the
stresses specified by \cref{Cts,Ctphi}. In a fully three-dimensional elastic
body, the Cauchy stress, $\vec{\sigma}$, is
\begin{equation}
 \vec{\sigma} = -p\mathbf{I} + 2\mathbf{F}\pdiff{W}{\mathbf{C}}\mathbf{F}^T\,, \label{C1}
\end{equation}
where $\mathbf{F}$ is the deformation gradient,
$\mathbf{C}=\mathbf{F}^T\mathbf{F}$ is the right Cauchy-Green tensor, $p$ is
the hydrostatic contribution to the stress associated with enforcing
incompressibility, and $W$ is the strain-energy function. For a detailed
account, we direct the interested reader to, for example, the work of
\cite{Holzapfel2010}. If we suppose that $W=W(I_1,I_4,I_6)$ where
\begin{subequations}
\begin{align}
 I_1 &= tr\mathbf{C}\,, \\
 I_4 &= \mathbf{a}^T\mathbf{C}\mathbf{a}\,, \\
 I_6 &= \mathbf{b}^T\mathbf{C}\mathbf{b}\,,
\end{align}
\end{subequations}
so that $I_4$ and $I_6$ represent the stretch of fibres that lie in the
directions $\mathbf{a}$ and $\mathbf{b}$, then
\begin{equation}
 \vec{\sigma} = -p\mathbf{I} + 2\mathbf{F}\left( \pdiff{W}{I_1}\mathbf{I} + \pdiff{W}{I_4}\mathbf{a}\otimes\mathbf{a} + \pdiff{W}{I_6}\mathbf{b}\otimes\mathbf{b}  \right)\mathbf{F}^T\,.\label{stressI4I6}
\end{equation}
Furthermore, if we select the basis used for our scleral model so that
$\mathbf{F}$ is diagonal with entries $\alpha_s,\alpha_{\phi},\alpha_n$,
define the fibre directions as in \cref{eq:fibre:a,eq:fibre:b,matfibre}, so
that $I_4=I_6$, and finally require $W(I_1,I_4,I_6)=W(I_1,I_6,I_4)$, then the
off-diagonal terms in $\mathbf{a}\otimes\mathbf{a}$ and
$\mathbf{b}\otimes\mathbf{b}$ cancel. Thus, the only non-zero components in
\cref{stressI4I6} are
\begin{subequations}
\begin{align}
 \sigma_{ss} &= -p + 2\alpha_s^2\pdiff{W}{I_1} + 4\alpha_s^2\sin^2\fibredirection\pdiff{W}{I_4}\,,\\
 \sigma_{\phi\phi} &= -p + 2\alpha_{\phi}^2\pdiff{W}{I_1} + 4\alpha_{\phi}^2\cos^2\fibredirection\pdiff{W}{I_4}\,,\\
 \sigma_{nn} &= -p + 2\alpha_n^2\pdiff{W}{I_1}\,.
\end{align}
\end{subequations}
The shell's thin geometry can be exploited as discussed in the context of
membranes in \cite{Haughton2001}. We apply the key results in our shell model
by working with resultant stresses of the form
\begin{subequations}
\begin{align}
 t_s&=\alpha_nH\sigma_{ss}\,, \\
 t_{\phi}&=\alpha_nH\sigma_{\phi\phi}
\end{align}
\end{subequations}
and setting $\sigma_{nn}=0$, often termed the `membrane assumption'. This is
akin to noting that the curved shell is so thin that load across the surface
due to the intraocular pressure is supported by in-shell tension, as opposed
to stress across the shell thickness. The membrane assumption enables the
elimination of the hydrostatic pressure, $p$, and our incompressibility
assumption, $\alpha_n = 1/\alpha_s\alpha_{\phi}$, gives principal in-shell
stress resultants of the form \cref{Cpst}.


\end{document}